\documentclass[lettersize,journal]{IEEEtran}
\usepackage{amsmath,amsfonts}
\usepackage{algorithmic}
\usepackage{algorithm}
\usepackage{array}
\usepackage[caption=false,font=normalsize,labelfont=sf,textfont=sf]{subfig}
\usepackage{textcomp}
\usepackage{stfloats}
\usepackage{url}
\usepackage{verbatim}
\usepackage{graphicx}
\usepackage[pdftex,dvipsnames]{xcolor}
\usepackage[colorinlistoftodos,prependcaption,textsize=tiny]{todonotes}
\usepackage{xargs}
\usepackage[utf8]{inputenc}
\usepackage[short]{optidef}
\usepackage{hyperref}
\usepackage{multirow}
\usepackage{amsthm}
\newtheorem{thm}{Theorem}[section]
\def\BibTeX{{\rm B\kern-.05em{\sc i\kern-.025em b}\kern-.08em
    T\kern-.1667em\lower.7ex\hbox{E}\kern-.125emX}}
\usepackage{balance}
\makeatletter
\def\endthebibliography{%
  \def\@noitemerr{\@latex@warning{Empty `thebibliography' environment}}%
  \endlist
}
\makeatother
\newcommand\twopartdef[4]
{
	\left\{
		\begin{array}{ll}
			#1 & \mbox{if } #2 \\
			#3 & \mbox{} #4
		\end{array}
	\right.
}
\newcommandx{\unsure}[2][1=]{\todo[linecolor=red,backgroundcolor=red!25,bordercolor=red,#1]{#2}}
\newcommandx{\change}[2][1=]{\todo[linecolor=blue,backgroundcolor=blue!25,bordercolor=blue,#1]{#2}}
\newcommandx{\info}[2][1=]{\todo[linecolor=OliveGreen,backgroundcolor=OliveGreen!25,bordercolor=OliveGreen,#1]{#2}}
\newcommandx{\improvement}[2][1=]{\todo[linecolor=Plum,backgroundcolor=Plum!25,bordercolor=Plum,#1]{#2}}
\newcommandx{\thiswillnotshow}[2][1=]{\todo[disable,#1]{#2}}

\begin{document}
\newcommand\myworries[1]{\textcolor{red}{#1}}
%\title{How to Use the IEEEtran \LaTeX \ Templates}
\title{A Supervisory Volt/VAR Control Scheme for Coordinating Voltage Regulators with Smart Inverters on a Distribution System}
\author{Valliappan Muthukaruppan,~\IEEEmembership{Student Member,~IEEE,} Yue Shi,~\IEEEmembership{Member,~IEEE,}

and Mesut E. Baran,~\IEEEmembership{Fellow,~IEEE}

\thanks{\textit{Corresponding Author}: Mesut Baran (\it{baran@ncsu.edu})}
\thanks{The authors are with Department of Electrical and Computer Engineering, North Carolina State University,
Raleigh, NC 27695 USA (email: vmuthuk2@ ncsu.edu, yueshi0430@gmail.com, and baran@ncsu.edu)}
}

% \markboth{Journal of \LaTeX\ Class Files,~Vol.~18, No.~9, September~2020}%
% {How to Use the IEEEtran \LaTeX \ Templates}

\maketitle

\begin{abstract}
  This paper focuses on the effective use of smart inverters for Volt/Var control (VVC) on a distribution system. New
  smart inverters offer Var support capability but for their effective use they need to be coordinated with existing
  Volt/Var schemes. A new VVC scheme is proposed to facilitate such coordination. The proposed scheme decomposes the
  problem into two levels. The first level uses Load Tap Changer (LTC) and Voltage Regulators (VRs) and coordinates
  their control with smart inverters to adjust the voltage level on the circuit to keep the voltages along the circuit
  within the desired range. The second level determines Var support needed from smart inverters to minimize the overall
  power loss in the circuit. The results of the supervisory control are sent to the devices which have their local
  controllers. To avoid frequent dispatch, smart inverters are supervised by shifting their Volt/Var characteristics as
  needed. This allows for the smart inverters to operate close to their optimal control while meeting the limited
  communication requirements on a distribution system. A case study using the IEEE 34 bus system shows the effectiveness
  of this supervisory control scheme compared to traditional volt/var schemes.
\end{abstract}

\begin{IEEEkeywords}
Volt/Var Optimization, Coordinated Control, Smart Inverters, Volt/Var Curve
\end{IEEEkeywords}

\section{Introduction}
Current Volt/VAR Control (VVC) schemes implemented by many utilities use legacy control devices such as capacitor banks,
load tap changers and voltage regulators. The primary goal is to keep the voltage across the feeder within the ANSI
limit \cite{zotero-512}. Recent increase in deployment of distributed energy resources (DERs) on distribution system
challenges effectiveness of the conventional VVC schemes \cite{sharma2020}. On the other hand, it was recognized that
the inverters used in most DERs can provide Var support. To encourage the use of this new capability, IEEE Std. 1547
has been revised and new operating modes with varying Volt-Var characteristics (VVar-C) have been developed for these
new smart inverters (SI) \cite{IEEE1547}. However, for the SIs to be effective in supporting VVC, their VVar-Cs need to
be chosen very carefully and need to be updated as system conditions change \cite{singhal2019}.

VVar-C based control for SIs is a popular local control framework which has been adopted by IEEE 1547 standard
\cite{IEEE1547} and adopted by Rule-21 in California \cite{Rule21}. In \cite{jahangiri2013}, the author points out that
the local control is effective in mitigating the voltage rise issue due to high PV penetration, but also points out the
oscillatory problem even for single SI with droop control scheme. In \cite{singhal2019}, the author addresses the low
steady state error (SSE) and stability/convergence issues associated with droop control by dynamically adapting the
droop parameters. In \cite{ding2018} a VVC scheme using legacy devices is proposed, and it incorporates SIs with fixed
VVar-C.

An alternative approach is to treat SIs as Var dispatchable sources and formulate the problem as Optimal Power Flow
(OPF) problem \cite{dallanese2014, farivar2013, robbins2016}. To address the computational challenges, two control loops
with different timescales are proposed in \cite{xu2019, jha2019}. In the slow update loop, the legacy devices are
dispatched, and in the fast update loop, SIs are dispatched by solving a separate Var optimization problem. In this
approach, the large number of discrete variables needed to represent VR operation increases the computational burden
significantly. To address this issue, we propose an efficient search algorithm for finding optimal tap positions of
VRs.

Another challenge in implementing an OPF based scheme is that sending the dispatching signals frequently which may not be
feasible with the limited communication infrastructure at distribution level \cite{manbachi2015, neal2011,
muthukaruppan2020a}.

This paper addresses these issues by introducing a new coordinated VVC scheme which adjusts the Volt-Var settings of SIs
periodically rather than sending Var commands directly. Also, a computationally efficient method for the VVC
method is proposed to facilitate its implementation in the field. This method also keeps legacy device operations low by
making use of the fast-acting capability of smart inverters. The main contribution of this paper are as follows:

\begin{itemize}

  \item VVC problem as a version of optimal power flow problem is NP-hard. There have been methods
  proposed based on convex relaxation of the original non-convex problem \cite{zhang2015, zheng2016} that can find global optimal solution under certain system
  conditions but they cannot handle discrete or integer variables associated with LTC and VR control
  \cite{bazrafshan2019}. On the other hand, linearizing the original power flow problem allows incorporation of LTC/VR
  control thereby allowing full coordination between the devices but the number of discrete variables introduced by
  LTC/VR control increases exponentially and may not be feasible for large systems \cite{jha2019}. To address these
  issues we propose a two-stage scheme. The first stage determines the set points for LTC, VR devices, and smart inverters by using an efficient search algorithm for VR/LTC tap settings. This provides a proper coordination between these devices and assures that voltages in the system will be within desired limits while keeping the number of tap changes low compared to conventional VVO schemes. The second stage checks if the Var dispatch for SIs can be further adjusted to decrease the power loss.
  
  \item To address the limited communication infrastructure available in field, we propose a new dispatching
  scheme for sending optimal Var commands to SIs through shifting of their volt/var curves. This scheme ensures that the smart inverters will continue to inject reactive power close to optimal value in between the dispatch intervals and provide additional fast response when there is intermittency from DERs.
  
\end{itemize}

The rest of the paper is organized as follows. In section~\ref{sec:coordinated_vvo}, we propose the coordination
strategy for dispatching LTC/VR and smart inverters simultaneously and in section~\ref{sub:VarLoop} we propose the
shifting curve strategy for smart inverter dispatch to avoid the frequent dispatch problem. The performance of proposed
control is then evaluated using a test system in section~\ref{sec:casestudy}.

\section{Supervisory VVO} % Coordinated VVO explanation
\label{sec:coordinated_vvo}
The application considered here is the VVC scheme to be implemented by a utility on a distribution system with SIs. The
proposed VVC scheme (i) coordinates the operation of SIs with the utility Volt-Var devices, and (ii) is computationally
efficient for easy adoption in practice.

A central VVC scheme will be most effective in this case to assure proper coordination between the utility VVC and SIs.
The dispatch speed for the controller is a critical issue at distribution level, as utilities generally utilize a
radio-mesh/cellular based communication infrastructure in distribution system. On the other hand, high intermittency
DERs necessitates that the dispatch needs to be as fast as possible. In this work, it is assumed that utility has a good
distribution SCADA which can collect data from the field for the control cycle.

\begin{figure}[htpb]
  \centering
  \includegraphics[width=0.4\textwidth]{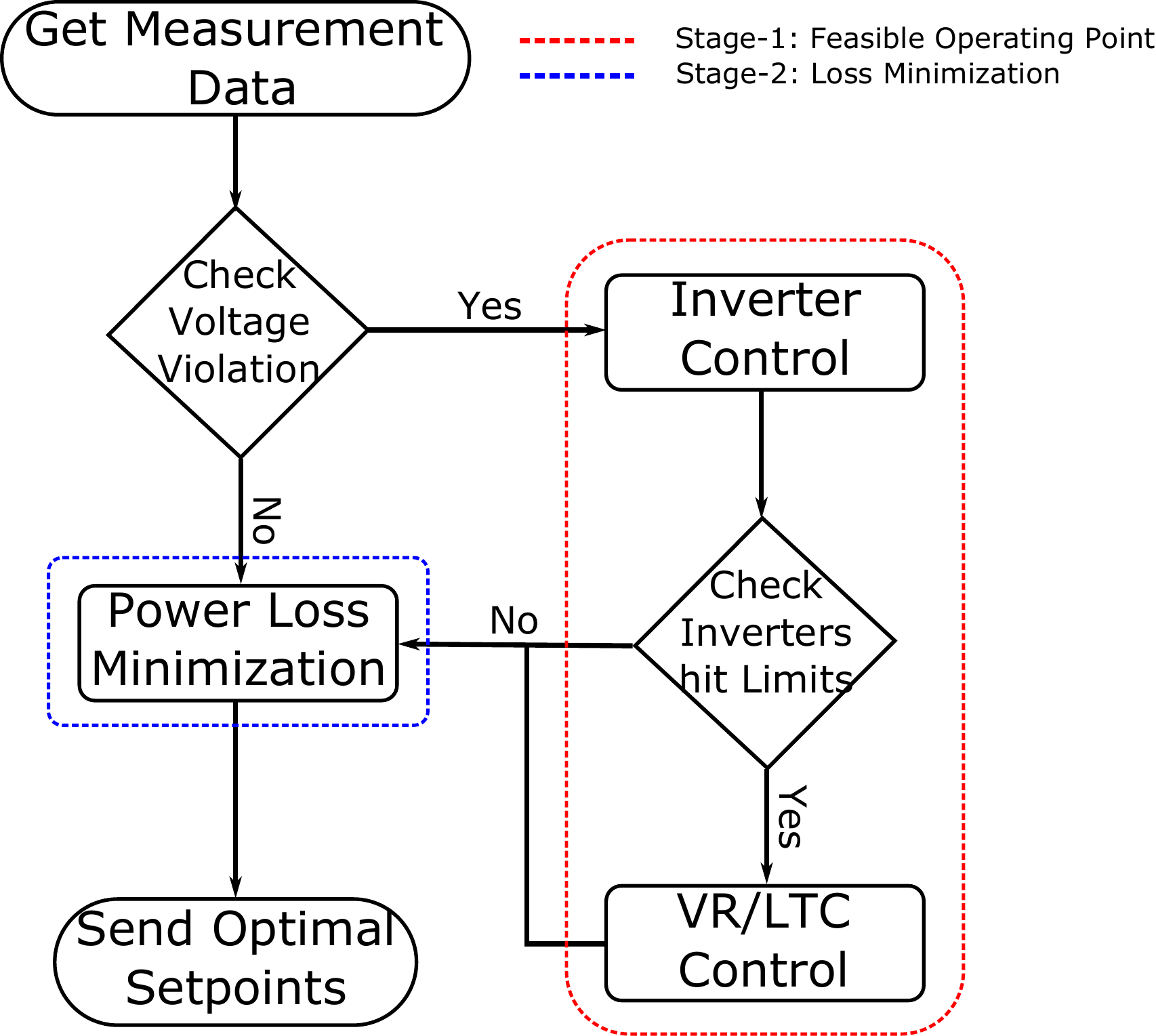}
  \caption{Flowchart of proposed algorithm}
  \label{fig:flowchart}
\end{figure}

The proposed scheme which is central will supervise both the conventional VVC devices and the SIs which can be
controlled by the utility for additional Var support. To reduce the computational burden, we adopt a two-stage
optimization scheme shown in fig. \ref{fig:flowchart}. As the figure indicates, stage-1 utilizes smart inverters and LTC/VRs to find a feasible operating point preparing for stage-2 which utilizes smart inverters to minimize the total power loss in the circuit. This approach allows SIs to respond to voltage violations first, thus avoiding unnecessary adjustments of VRs. These two stages are elaborated below.

\subsection{Stage-1: Feasible Operating Point}\label{subsec:stage1}
The goal of stage-1 is to bring the voltages well within the desired limits. To minimize the use of legacy devices which can be subject to mechanical wear and tear
with frequent operation, we utilize SIs first to bring the voltage within ANSI limits ($v\in[0.95, 1.05]$ p.u.) and
only if the inverters hit their power limits or voltage violations still exist after using inverters we use LTC/VR to
regulate the voltage.

\subsubsection{Inverter Control}\label{subsec:invCtrl}
The objective of inverter control in stage-1 is to flatten the voltage across the system. Let $g(v, q)$ denote the power
flow constraints which is a set of equations $\forall i\in\mathcal{N}$ given by (\ref{eqn:2}) and (\ref{eqn:3}) where
$\mathcal{N}$ is the set of system nodes.
\begin{align}
  p_{i}^{g} - p_{i}^{l} &= v_{i}\sum_{j\in\mathcal{N}} v_{j}(G_{ij}\cos\theta_{ij} + B_{ij}\sin\theta_{ij})\label{eqn:2}\\
  q_{i}^{g} - q_{i}^{l} &= v_{i}\sum_{j\in\mathcal{N}} v_{j}(G_{ij}\sin\theta_{ij} - B_{ij}\cos\theta_{ij})\label{eqn:3}
\end{align}

Here, $p_i^g$ is the real power generation from PV at node $i$, $p_i^l$ is the real power consumption at node $i$,
$q_{i}^g$ is the reactive power injection from smart inverters, and $q_i^l$ is the reactive power consumption.
$\theta_{ij} = \theta_i - \theta_j$ is the voltage difference between node $i$ and $j$, $G_{ij} + jB_{ij} = Y_{ij}$ is
the element of Y-bus matrix. The mathematical formulation of this problem is given in (\ref{eq:4}) which is solved using
a gradient-descent approach.
\begin{mini}[4]
  {q^g}{\sum_{i\in\mathcal{N}}\omega_i(v_i - v_{ref})^2}{\label{eq:4}}{}
  \addConstraint{}{g(\mathbf{v}, \mathbf{q})=0}{}
  \addConstraint{}{v_{min}\le v_i}{\le v_{max}}{\quad\forall i\in\mathcal{N}}
  \addConstraint{}{q^g_{min}\le q_i^g}{\le q^g_{max}}{\quad\forall i\in\mathcal{N}}
  \addConstraint{}{\text{where,}}{\twopartdef {\omega_{i}=0}{v_{i} \in [0.95, 1.05]} {\omega_{i}=1}{\text{otherwise}}}
\end{mini}

Here, $v_{ref}$ is the desired voltage profile through out the system, generally fixed at 1.0 p.u. $\mathbf{v}$ is the
vector of voltage magnitudes and $\mathbf{q}$ is the vector of reactive power injections.

\subsubsection{VR/LTC Control}\label{subsec:ltcvr}
As indicated earlier LTC/VR control is only utilized when the inverters are unable to regulate the voltage. The LTC and
VR tap positions are discrete; these devices usually have 33 taps (including the zero-tap position) and each tap
corresponds to a 0.00625 p.u. voltage change. The LTC at substation is typically three-phase gang-operated, while the
VRs are usually controlled on a phase basis to be able to adjust voltage on each phase independently.

Due to discrete tap control, an exhaustive search for feasible tap settings can be computationally challenging. With
R single phase controlled regulators and K three phase controlled regulators, the total search space will be
$33^{3R+K}$. For example, a system with one LTC and one VR, total possible tap combinations will be
33x33x33x33=1,185,921. To reduce the search, a search method based on \cite{ozdemir2016} has been developed. The method
is based on the observation that most of the time tap adjustments needed are small. Hence, instead of searching all
possible 33 taps, the method reduces the feasible search space to at most 2 to 3 taps up or down, $\Delta Tap = \{0, \pm
1, \pm 2, \pm 3\}$.

\begin{figure}[htpb]
  \centering
  \includegraphics[width=0.5\textwidth]{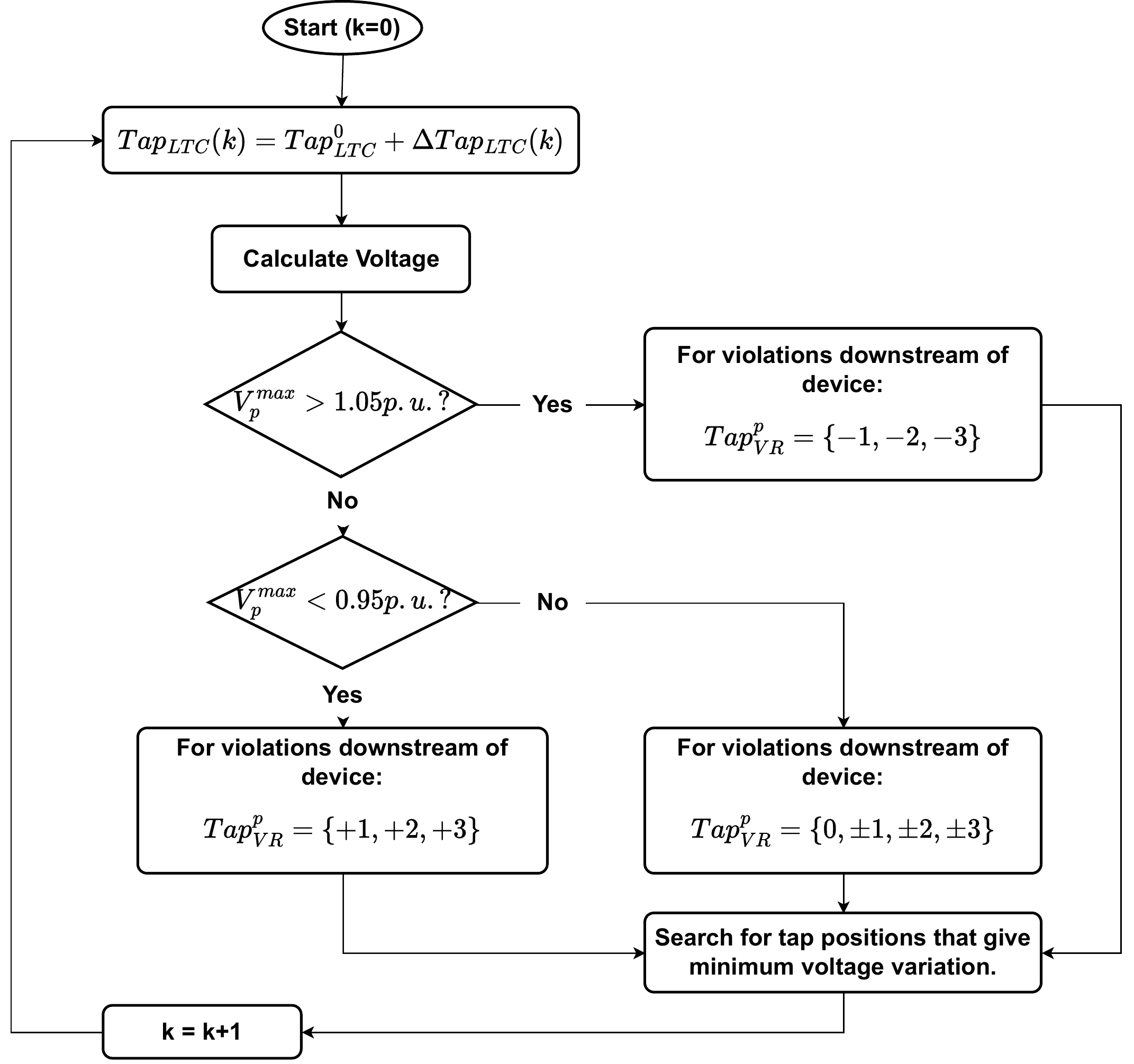}
  \caption{Flowchart of VR/LTC Module}
  \label{fig:flowchart-VR}
\end{figure}

In this work we further reduce this search space by applying additional rules as given in fig. \ref{fig:flowchart-VR}.
As the figure shows, voltages are calculated by varying the voltage control device closest to the substation (e.g. LTC)
with tap adjustment $Tap_{LTC}(k) \in \Delta Tap$, while keeping tap position constant for downstream voltage control
devices. If over-voltage violations occur, only down taps of downstream devices are searched. If under-voltage
violations occur, only up taps of downstream devices are searched. In fig. \ref{fig:flowchart-VR}, p represents the
corresponding phase of voltage regulator. We use a backward-forward three-phase distribution power flow solver to
determine the voltages for different tap combinations during the search. This approach reduces the overall search space
to $3^{3R} + K$ on average and $7^{3R} + 3^{K}$ on worst case.

Since there will be many tap settings that will bring the voltages within the limits, a criterion is needed to
determine the preferred settings for the legacy devices. In our scheme, keeping the voltage profile as flat as possible
is selected as the criteria, as this is the preferred voltage profile for a feeder. Hence, in the search, the feasible
VR settings are ranked using the voltage variance criterion as shown in (\ref{eqn:var}), and after the search the VR tap
settings with best voltage profile is selected as the new tap settings for VRs. The $V_{ref}$ is generally chosen as 1.0
p.u. and having a flat voltage profile can lead to lower power loss and power consumption in the circuit through
Conservation Voltage Reduction (CVR) \cite{jha2019}. 
\begin{equation} \label{eqn:var}
  V_{var} = \sum_{i\in \mathcal{N}}^{} \left(V_i - V_{ref}\right)^2
\end{equation}

\subsection{Stage-2: Loss Minimization}%
\label{sub:VarLoop}
The objective in this stage is to optimally dispatch the reactive power injections of SIs to minimize
the total power loss in the circuit while keeping the voltages within limits. The mathematical formulation of this
problem is given in (\ref{eq:6}) which is also solved using a gradient-descent approach:
\begin{mini}[4]
  {q^g}{\sum_{i\neq j}G_{ij} \left[ v_{i}^{2} + v_{j}^2 - 2v_{i}v_{j}\cos\theta_{ij} \right]}{\label{eq:6}}{}
  \addConstraint{}{g(\mathbf{v}, \mathbf{q})=0}{}
  \addConstraint{}{v_{min}\le v_i}{\le v_{max}}{\quad\forall i\in\mathcal{N}}
  \addConstraint{}{q^g_{min}\le q_i^g}{\le q^g_{max}}{\quad\forall i\in\mathcal{N}}
  \addConstraint{}{\text{where,}}{\twopartdef {\omega_{i}=0}{v_{i} \in [0.95, 1.05]} {\omega_{i}=1}{\text{otherwise}}}
\end{mini}

\subsection{Dispatching of Smart Inverter}
\label{sub:dispatch}
The set points obtained by VVO need to be dispatched to the SIs on the system. Due to limited communication infrastructure and bandwidth utilities are unable to dispatch optimal var commands to inverters frequently. On the other hand, relying completely on local control approach using VVar-C can impact the overall VVO since they follow a fixed VVar-C. The new dispatch strategy we propose for SIs uses the Var settings from VVO and determines how to adjust VVar-C of SIs so that they will provide Var support indicated by VVO during the dispatch period. 

\begin{figure}[htpb]
  \centering
  \includegraphics[width=0.48\textwidth]{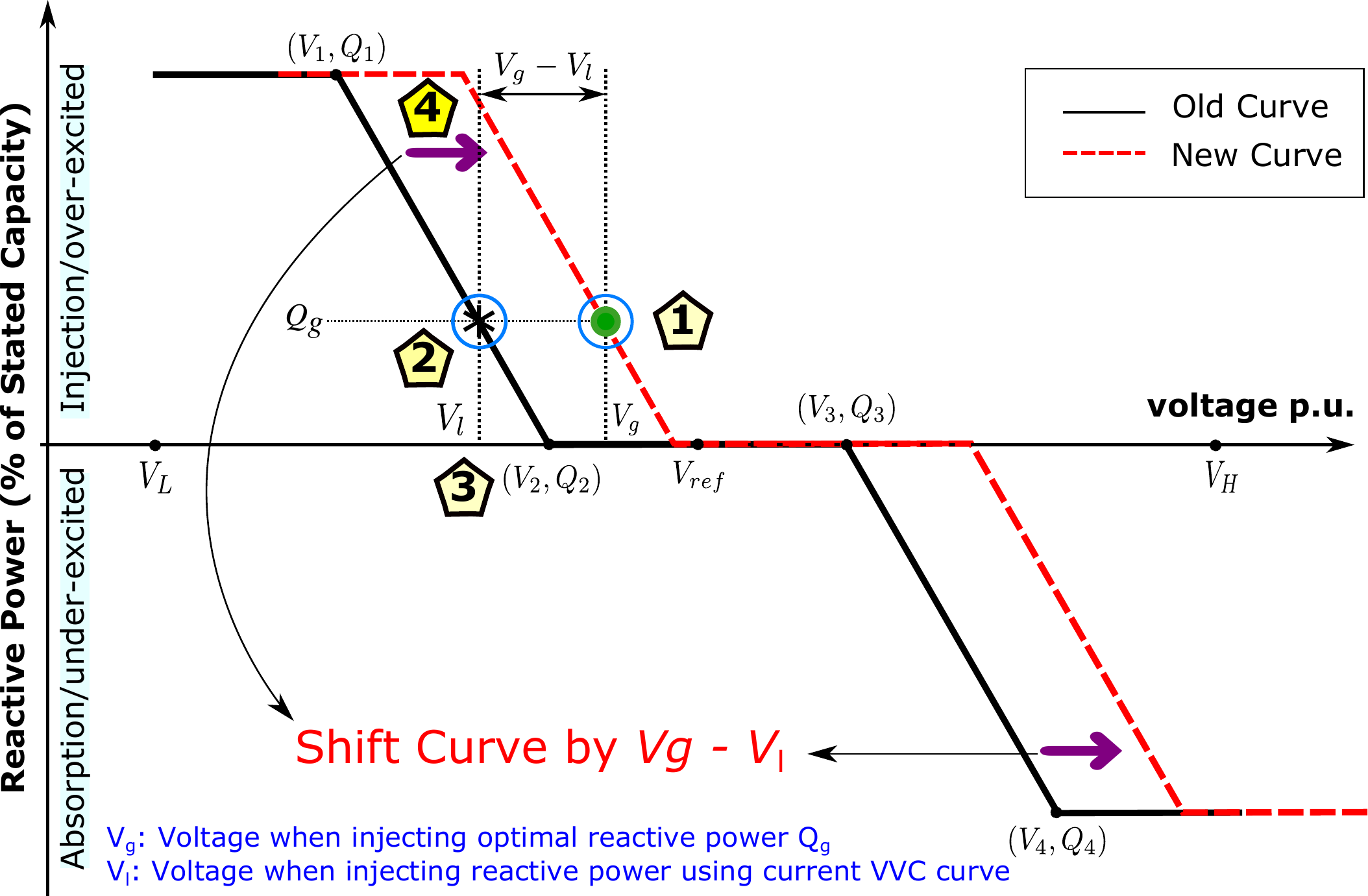}
  \caption{VVar-C Shifting Strategy}
  \label{fig:curveshift}
\end{figure}

The VVar-C is generally characterized by a slope $\Delta Q$ and a reference point $V_{ref}$ which is realized using four
points $(V_1, Q_1), (V_2, Q_2), (V_3, Q_3)$ and $(V_4, Q_4)$ as shown in fig-\ref{fig:curveshift}. Where $Q_1$, $Q_4$
correspond to $Q_{lim}$ of the inverter and $Q_2, Q_3$ are generally 0.

In \cite{zhou2021a}, the author disseminates the engineering of local control and provides a framework for perceiving
the dynamical nature of VVar-C as a distributed optimization problem that strives to minimize the deviation of system
voltage from nominal value ($v_{ref}$) while simultaneously minimizing the reactive power provisioning required at each
inverter.

In \cite{zhou2021a} the author proves that the equilibrium point of this curve is a unique point ($v^*$,  $q^*$) which
satisfies the power flow equations $g(v^*, q^*)$ and the VVar-C curve $q^* = f(v^*)$. We use this theorem to shift
the VVar-C from current curve to new curve based on the optimal solution ($v^*, q^*$) obtained from the centralized
optimization problem. We prove in theorem-\ref{thm:curveshifting} that by shifting the curve, new equilibrium point of
the curve after the dynamics will be the optimal set point $(v^*, q^*)$.

\begin{thm}\label{thm:curveshifting}
Given an optimal power flow solution ($v_g, q_g$) for any operating point, by shifting the existing volt/var curves $q
= f_1(v - v_{ref})$ where  $f_1\colon \mathbf{R}^n \to \mathbf{\Omega}$ is collection of volt/var function of all
inverters in the system, along the voltage axis by  $v^g - v^l$ where $v^l = f_1^{-1}(q^g)$ is voltage corresponding to
$q^g$ on existing volt/var curve $f_1$ will result in a new set of volt/var curves $f_2$ whose equilibrium point will be
($v_g, q_g$). 
\end{thm}
\vspace{-10pt}
\begin{equation}\label{eqn:thm}
\begin{aligned}
  &&q_g &= f_2(v_g - v_{ref})\\
  &\text{where, } &f_2(v - v_{ref}) &= f_1(v - v_{ref} + v_g - v_l) 
\end{aligned}
\end{equation}

Proof of theorem-\ref{thm:curveshifting} is provided in Appendix. Suppose say $q_g$ is the optimal solution of
a inverter to our centralized problem and $v_g$ be the corresponding voltage at the inverter which implies $g(v_g, q_g)
= 0$ since it is also a feasible solution. Now, using the current VVar-C at the inverter we find the voltage $v_l$ that
will correspond to the optimal var $q_g$. Note that ($v_l, q_g$) is not a equilibrium point of the VVar-C since it does
not satisfy the power flow constraints. By shifting the four set points of the curve by $v_g - v_l$, the new equilibrium
point will in fact fall at ($v_g, q_g$) according to theorem-\ref{thm:curveshifting}. The direction of shift will be
intrinsically handled by the sign of the difference $v_g - v_l$. The dispatching algorithm is explained in
algorithm-\ref{alg:shift} and highlighted in fig. \ref{fig:curveshift}.

 \begin{algorithm}[htbp]
 \caption{VVar-C Shifting Algorithm}
 \label{alg:shift}
 \begin{algorithmic}[1]
 \renewcommand{\algorithmicrequire}{\textbf{Input:}}
 \renewcommand{\algorithmicensure}{\textbf{Output:}}
 \REQUIRE Old VVar-C - $\hat{f_i}(v) \quad \forall i \in \mathcal{N}_{SI}$
 \ENSURE New VVar-C - $f_i(v) \quad \forall i \in \mathcal{N}_{SI}$
 % \\ \textit{Initialization} : MA model parameters $\boldsymbol\theta$ and $\boldsymbol\mu$
 \STATE Run the centralized optimization problem to obtain the individual var commands $q^g_i$ and the corresponding
 voltage  $v^g_i$  $\forall i \in \mathcal{N}_{SI}$.
 \STATE Obtain the voltage $v^l_i$ corresponding to $q^g_i$ using the old curve $\hat{f}_i(v^l_i)$. 
 \STATE Shift the curve by $v^g_i - v^l_i$ which leads to new curve $\hat{f}_i(v^l_i + v^g_i - v^l_i) = f_i(v^g_i)$.
 \STATE With this new curve $f_i(v^g_i)$, the equilibrium point for current operating condition will be $(q^g_i,
 v^g_i)$. Dispatch the new curves $f_i(v)$ to the inverters.
 \end{algorithmic}
 \end{algorithm}

With this framework only the shift in curve $v_g - v_l$ needs to be dispatched to each individual inverter instead of
all the four setpoints thereby reducing the communication overhead required. Also, when the  $q_g$ is very small (say
below 0.1 p.u) then the new curve is not dispatched and that particular inverter is left at old curve to further reduce
the communication burden. It is important to note that with this shifting strategy only the $v_{ref}$ of the inverters
are modified but the slope of the VVar-C remains same thereby avoiding any stability issues in the control.

\section{Case Study}\label{sec:casestudy}
\begin{figure}[htpb]
  \centering
  \includegraphics[width=0.48\textwidth]{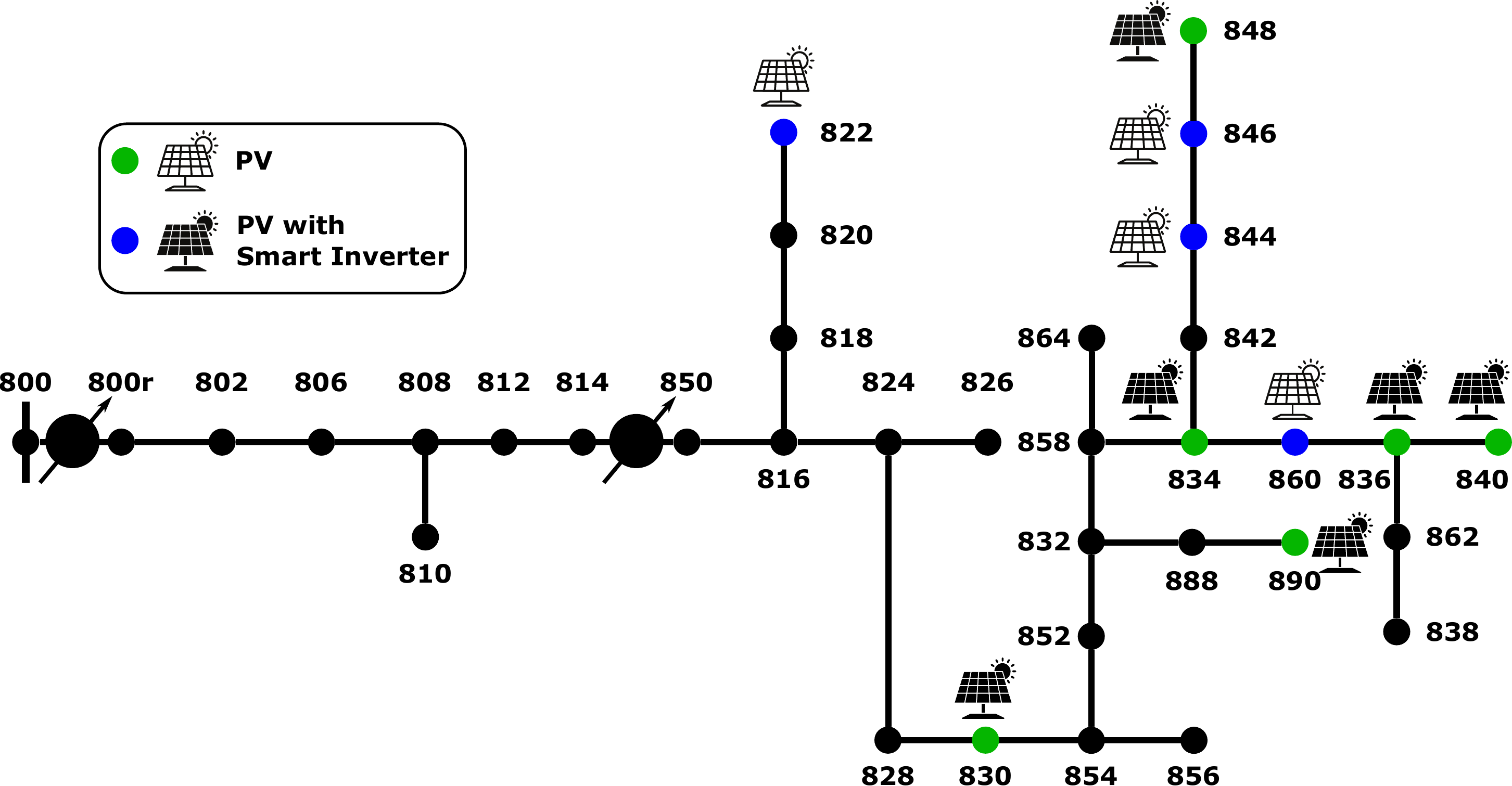}
  \caption{Test System with Location of PV and Smart Inverters.}
  \label{fig:testsystem}
\end{figure}

\begin{figure}[htpb]
  \centering
  \includegraphics[width=0.5\textwidth]{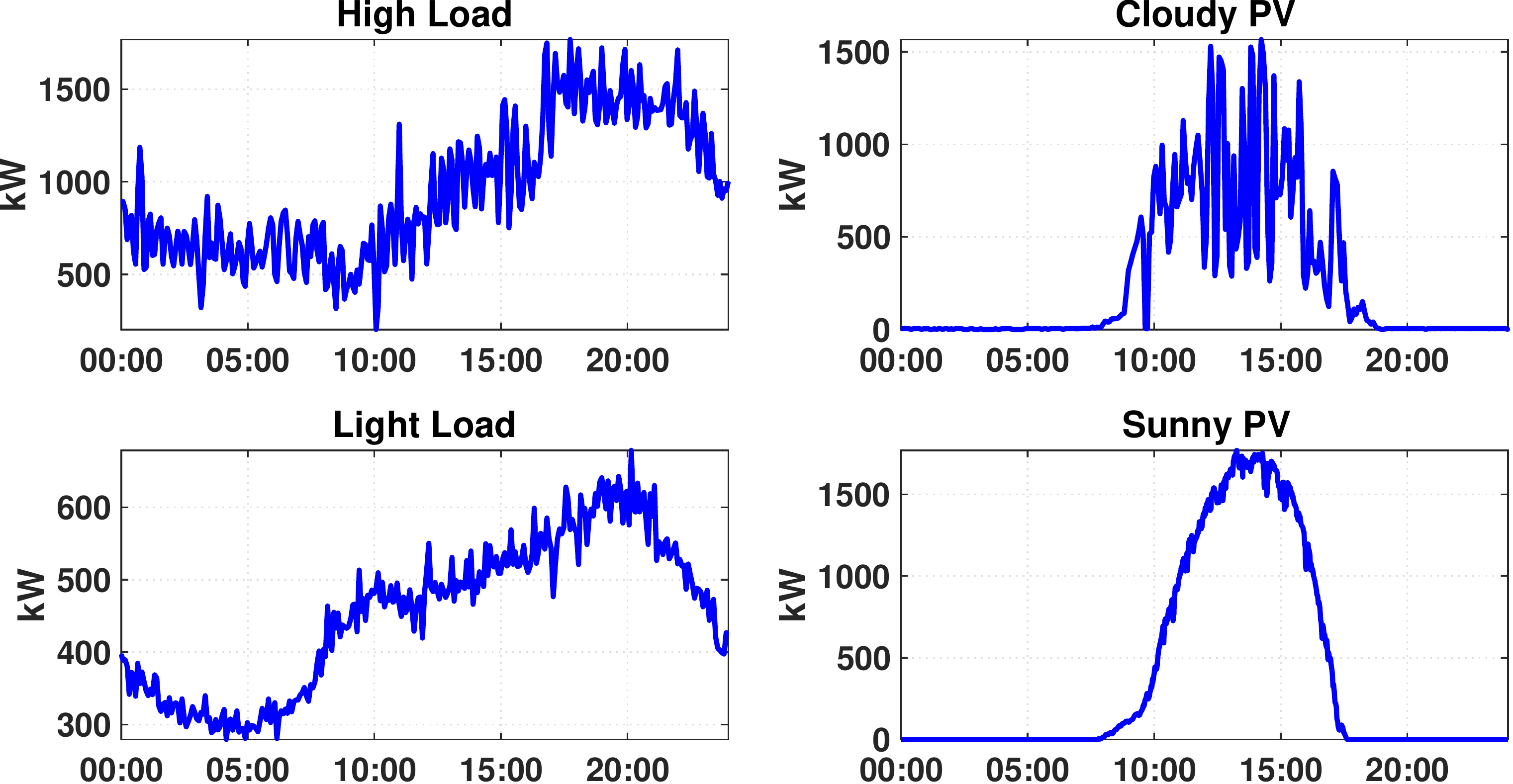}
  \caption{Load and PV profiles used in the case study.}
  \label{fig:profiles}
\end{figure}

A modified version of the IEEE 34 node system is utilized for simulating the test cases as shown in fig.
\ref{fig:testsystem}. This test feeder is based upon an actual feeder in rural Arizona. The transformer between nodes
832 and 888 is removed from the original feeder. A three-phase controlled LTC is installed at the substation and $VR_1$
is retained at its original location while  $VR_2$ between nodes 852 and 832 is removed. The  $VR_1$ and LTC have
$\pm 16$ tap position range with $\pm 10\%$ of maximum voltage change. The three phases of the VRs are controlled
separately. All shunt capacitors are removed from the original system. There are 20 nodes with load connected to
them in the original 34 node system. We connect 10 nodes with PV system and of these 6 have smart inverter 
VVar-C capability as highlighted in the figure \ref{fig:testsystem}. The 24 hour total  load and PV profiles
simulated are shown in fig. \ref{fig:profiles}. We evaluate the algorithm against four different operating
conditions consisting high load condition in summer, light load condition in winter, cloudy PV day with lot of
intermittency in PV output and a clear sunny day with peak PV output.

\subsection{Test Cases}
To assess the performance of the proposed scheme three VVC schemes are considered:

\begin{itemize}
  \item \textbf{Case-1}: LTC and VR operate under local control. PV system are present in the circuit but they donot
  have any smart inverter capability and hence do not provide any VAR support.
  \item \textbf{Case-2}: LTC and VR act under local control. PV system have smart inverter capability and act under
  local control with a fixed VVar-C curve.
  \item \textbf{Case-3}: Proposed coordinated VVO scheme with coordination between legacy devices and smart inverters.
  Also the proposed dispatch scheme is used to dispatch VVar-C curves.
\end{itemize}

\subsubsection{Case-1}\label{sec:case1}
In this case, only conventional devices participate in the volt/var control. The control is purely based on local
measurements \cite{short2004}. The LTC and VR control settings are highlighted in table-\ref{tab:case1}.
\begin{table}[htbp]
\centering
\caption{Set points for Legacy Devices}
\label{tab:case1}
  \begin{tabular}{l|c|c}
    \hline
    \multicolumn{1}{c|}{\textbf{Setting}} & \textbf{VR}   & \textbf{LTC}\\ \hline \hline
    $V_{set}$                              & 120V         & 122V        \\ 
    Bandwidth                              & 2V           & 2V          \\ 
    Time Delay                             & 60s          & 30s         \\ 
    Max tap Change                         & 1            & 1           \\ \hline
  \end{tabular}
\end{table}

Even though PV inverters are shown in fig. \ref{fig:testsystem} they do not participate in the Volt/Var Control. The
objective of setting up this case is to show that legacy devices are not capable of restricting voltage violations in presence of high PV.

\subsubsection{Case-2}\label{sec:case2}
\begin{figure}[htpb]
  \centering
  \includegraphics[width=0.35\textwidth]{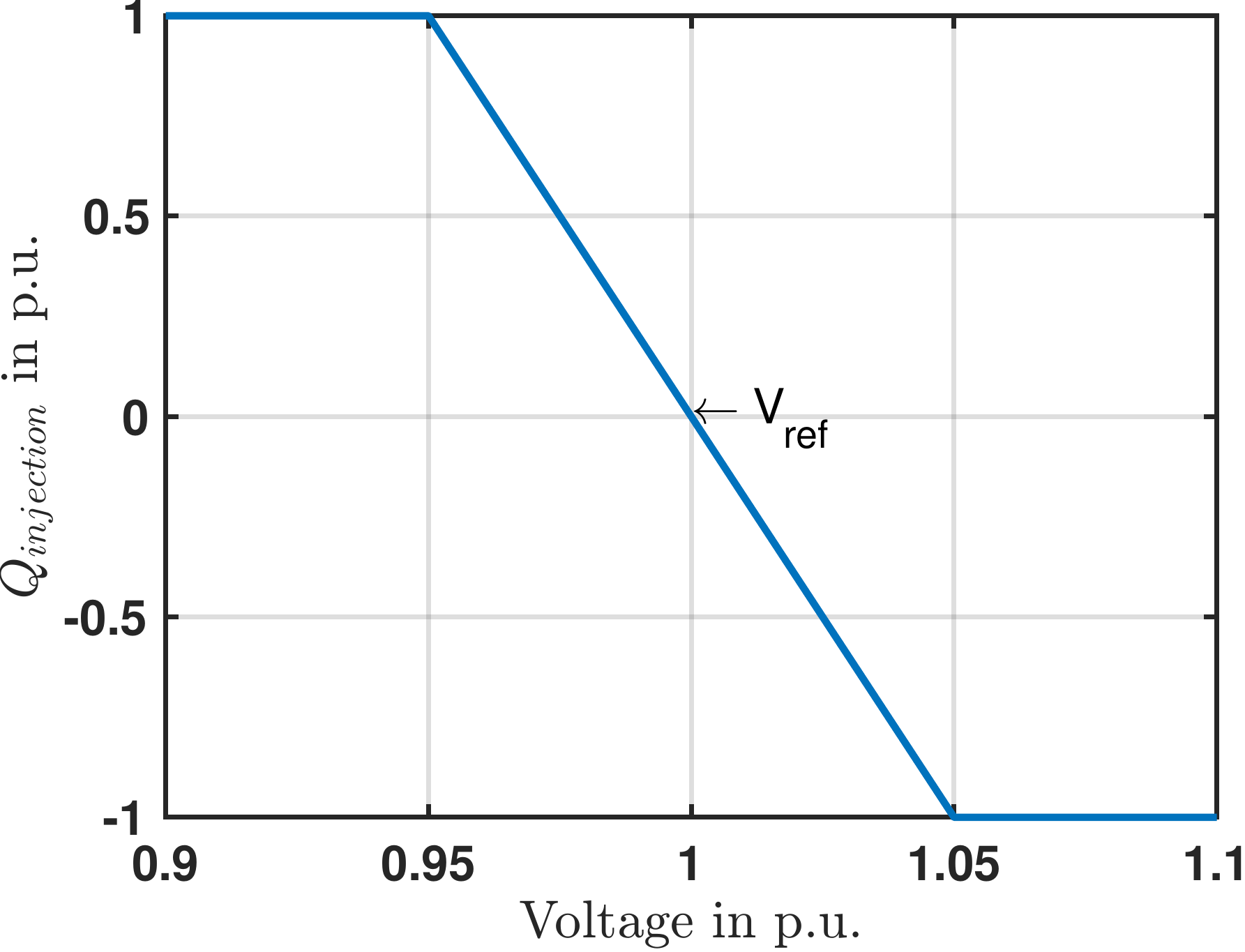}
  \caption{Default VVar-C for Smart Inverters.}
  \label{fig:vvarc}
\end{figure}

In this case, apart from the LTC and VR, the smart inverters in the system also participate
in the volt/var control. All devices operate based on local measurements and the control strategy of LTC and VR are same
as case-1. All the smart inverters here will have a fixed volt/var curve as shown in fig. \ref{fig:vvarc} and operate based on
the local voltage measurements at the inverter terminals.

\subsubsection{Case-3}\label{sec:case3}
Case-3 is the proposed coordinated VVO with curve shifting dispatch scheme. All inverters are initialized with the fixed
curve as shown in fig. \ref{fig:vvarc} but will be shifted based on the optimal var injections obtained from the var
optimization module every 5 minutes (assumed dispatch period) and it dispatches tap change commands to legacy devices and the shifted VVar-C to SIs. The smart inverters will use the new VVar-C and the local voltage measurements to inject necessary reactive power during the sub 5-minute interval. All SIs are initialized with the fixed curve as shown in fig. \ref{fig:vvarc}.

As explained in sec-\ref{sub:VarLoop}, to avoid frequent shifting of VVar-C a threshold of 10\% total var is introduced
in the shifting algorithm. Whenever the optimal var injection $q_g$ for an inverter is less than 10\% of total var limit
of the inverter, the VVar-C is retained at old curve and no shifting is applied to that inverter.

\subsection{Test Results}\label{sec:results}
We use various metrics to evaluate the performance of the proposed VVO scheme. During the 24 hour simulation: 
\begin{enumerate}
  \item $N_{UV}$ is the number of under voltage violations.
  \item $N_{OV}$ is the number of over voltage violations.
  \item Loss (kWh) is the total loss realized in the network.
  \item $N_{LTC}$ is the number of LTC tap operations.
  \item $N_{VR}$ is the number of VR tap operations. 
  \item $N_{total} = N_{LTC} + N_{VR}$ is the total number of legacy device operation. 
  \item $V_{max}$ is the maximum voltage recorded in the system.
  \item $V_{min}$ is the minimum voltage recorded in the system.
\end{enumerate}

\begin{table*}[htbp]
\centering
\caption{Case Study Comparison under Different Operating Conditions}
\label{tab:caseComp}
\begin{tabular}{l|c|c|c|c|c|c||c|c|c|c|c|c}
\hline
\multicolumn{1}{c|}{\multirow{3}{*}{\textbf{Metrics}}} & \multicolumn{6}{c||}{\textbf{High Load}}
& \multicolumn{6}{c}{\textbf{Light Load}}                                                                   \\
\cline{2-13}
\multicolumn{1}{c|}{}                                  & \multicolumn{3}{c|}{\textbf{Cloudy PV}}
& \multicolumn{3}{c||}{\textbf{Sunny PV}}               & \multicolumn{3}{c|}{\textbf{Cloudy PV}}
& \multicolumn{3}{c}{\textbf{Sunny PV}}               \\ \cline{2-13}
\multicolumn{1}{c|}{}                                  & \textbf{Case-1} & \textbf{Case-2} & \textbf{Case-3}
& \textbf{Case-1} & \textbf{Case-2} & \textbf{Case-3} & \textbf{Case-1} & \textbf{Case-2} & \textbf{Case-3}
& \textbf{Case-1} & \textbf{Case-2} & \textbf{Case-3} \\ \hline \hline
$N_{OV}$                                              & 2               & \bf0               & \bf0
& 0               & 0               & 0               & 39              & \bf0               & \bf0               & 153
& \bf0               & \bf0               \\
$N_{UV}$                                              & 228             & 74              & \bf0               & 219
& 75              & \bf0               & 0               & 0               & 0               & 0               & 0               & 0               \\
Loss (kWh)                                            & 1373.02         & 1510.58         & \bf1189.07         & 1358.04
& 1542.39         & \bf1233.23         & \bf316.97          & 575.97          & 351.44          & \bf466.41          & 730.56          & 512.6           \\
$N_{LTC}$                                             & \bf2               & \bf2               & 21              & \bf2
& \bf2               & 8               & 3               & 3               & \bf1
& 3               & 3               & \bf2               \\
$N_{VR}$                                              & 629             & 284             & \bf106             & 253
& 68              & \bf47              & 372             & 106             & \bf12              & 56              & 27
& \bf21              \\
$N_{total}$                                           & 631             & 286             & \bf127             & 255
& 70              & \bf55              & 375             & 109             & \bf13              & 59              & 30
& \bf23              \\
$V_{max}$                                             & 1.056           & 1.025           & 1.05            & 1.05            & 1.025           & 1.05            & 1.062           & 1.03            & 1.043           & 1.058           & 1.027           & 1.048           \\
$V_{min}$                                             & 0.909           & 0.9375          & 0.956           & 0.909
& 0.9375          & 0.951           & 0.952           & 0.963           & 0.964           & 0.954           & 0.963
& 0.956   \\ \hline       
\end{tabular}
\end{table*}

The simultaion results are summarized in table-\ref{tab:caseComp} for the four different operating conditions as per fig.
\ref{fig:profiles}. The key insights are summarized below:

Under all operating conditions the proposed VVO scheme (case-3) eliminates over voltage and under voltage issues
indicated by the $N_{UV}$ and  $N_{OV}$ values. Case-1 performs poorly as it can neither address the under voltage
issues during peak load condition nor the over voltage issues during peak PV condition which indicates that the legacy
devices are incapable of addressing voltage issues in presence of PV. Even though case-2 with fixed VVar-C is capable of
drastically reducing these violations, it still cannot completely eliminate the violations. But case-2 still performs
really well in eliminating the over-voltage issues caused by PV as highlighted in the results of light load operating
conditions.

Another major advantage of the proposed VVO scheme is the significant reduction in operation of legacy devices. In
case-1 the overall operation of LTC and VR is significantly high since in the absence of smart inverter capability, all
the intermittency in PV is handled by the legacy devices. The VR operations are higher than LTC due to the time delay between
them. On the other hand, case-2 manages to reduce the overall tap operations from case-1 but still the $N_{total}$ is
significantly high due to lack of coordination between the inverter control and legacy device control. Moreover, even
with such high tap operations both the cases perform poorly in eliminating the voltage issues. The proposed scheme
manages to considerably reduce the overall number of tap operations while completely eliminating any voltage violations
in the circuit. The reduction is extensively highlighted during cloudy PV days where proposed scheme reduces operation
by about 90\% compared to case-1 and 70\% compared to case-2.

Since, in the proposed scheme we optimize the var injections for total power loss minimization, we can observe
a considerable reduction in the total power loss realized during the 24 hour simulation. During high load condition we
can see that the proposed scheme has the lowest loss compared to case-1 and case-2. Case-2 has particularly high loss
because local control approach is known to increase the power loss due to lack of coordination in control between
the inverters. This is also evident from case-1 loss which is lower than case-2 in the absence of inverter control.
During light load condition the power loss of case-1 is lower than proposed scheme because the overall reactive load in
the circuit is low and with inverter control in proposed scheme the loss is higher due to var injections from the smart
inverters.
\begin{figure}[htpb]
  \centering
  \includegraphics[width=0.48\textwidth]{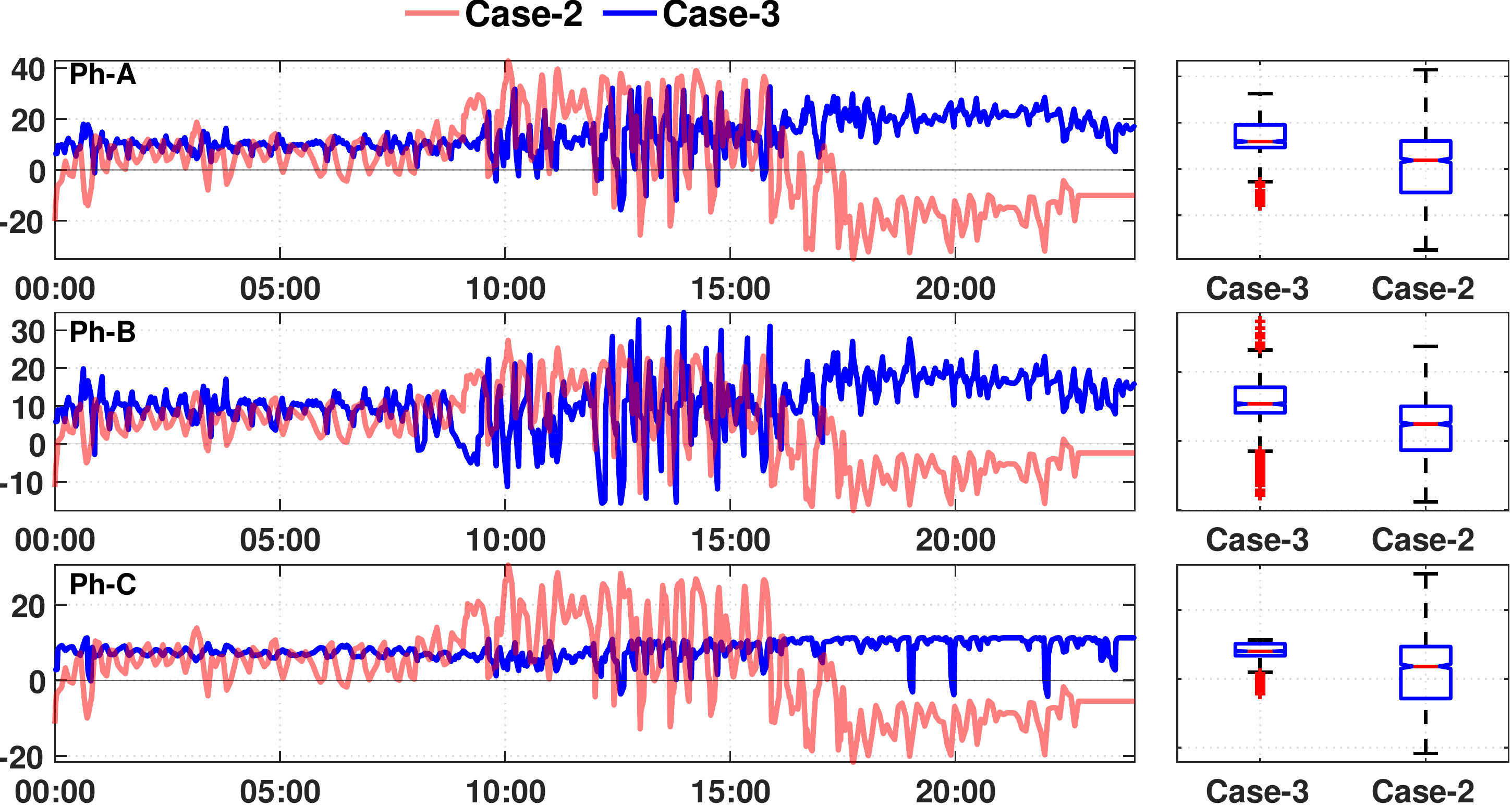}
  \caption{Comparison of Reactive Power Injections from Inverter at node 840 for High Load Cloudy PV condition.}
  \label{fig:qplot}
\end{figure}

Figure \ref{fig:qplot} shows the reactive power injections from inverter at node 840 during high load cloudy PV
condition between case-2 and case-3. In all three phases of the inverter case-3 manages to use very low var to regulate
the voltage which leads to much less power loss as seen in table-\ref{tab:caseComp}. This shows the advantage of
shifting the curves compared to case-2. 
\begin{figure}[htpb]
  \centering
  \includegraphics[width=0.48\textwidth]{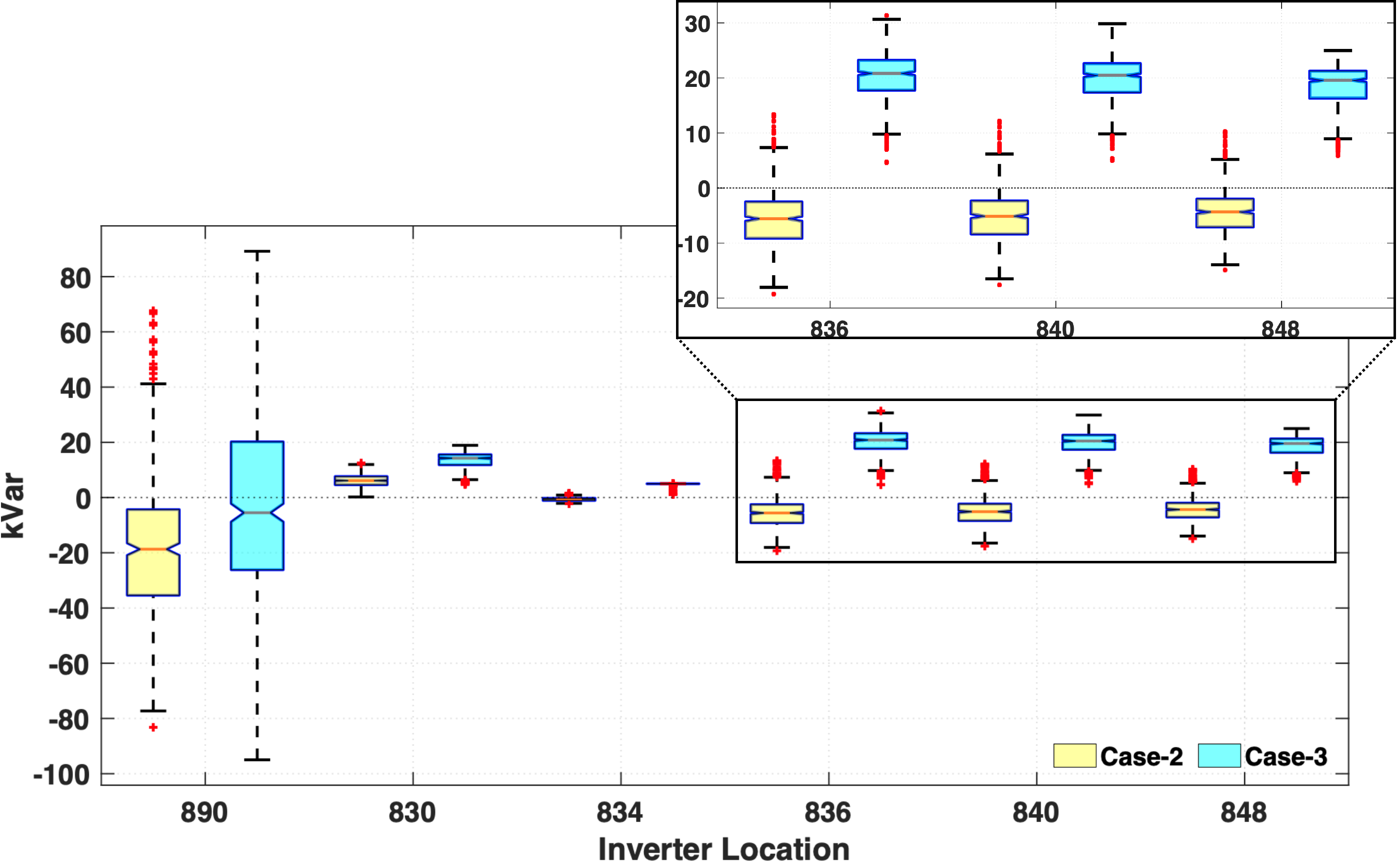}
  \caption{Distribution of Reactive Power Injections from 16:00-24:00 for High Load Cloudy PV Condition.}
  \label{fig:qboxplot}
\end{figure}

Further investigation shows that by shifting the curves we are managing the reactive power requirements from the
inverters strategically depending on size and location of the inverters on the circuit. As highlighted in fig.
\ref{fig:qboxplot} the reactive power from the inverter at node 890 which has a large PV system in the middle of the
circuit is utilized as conventional VVar-C device injecting reactive power at high load condition
while inverters towards the end at node 836, 840, and 848 with comparatively lower size are operated as reactive power
absorbing devices. While in case-2 the reactive power requirements are equally distributed among the inverters which all
act as reactive power injecting devices during high load condition. 

\begin{figure}[htpb]
  \centering
  \includegraphics[width=0.48\textwidth]{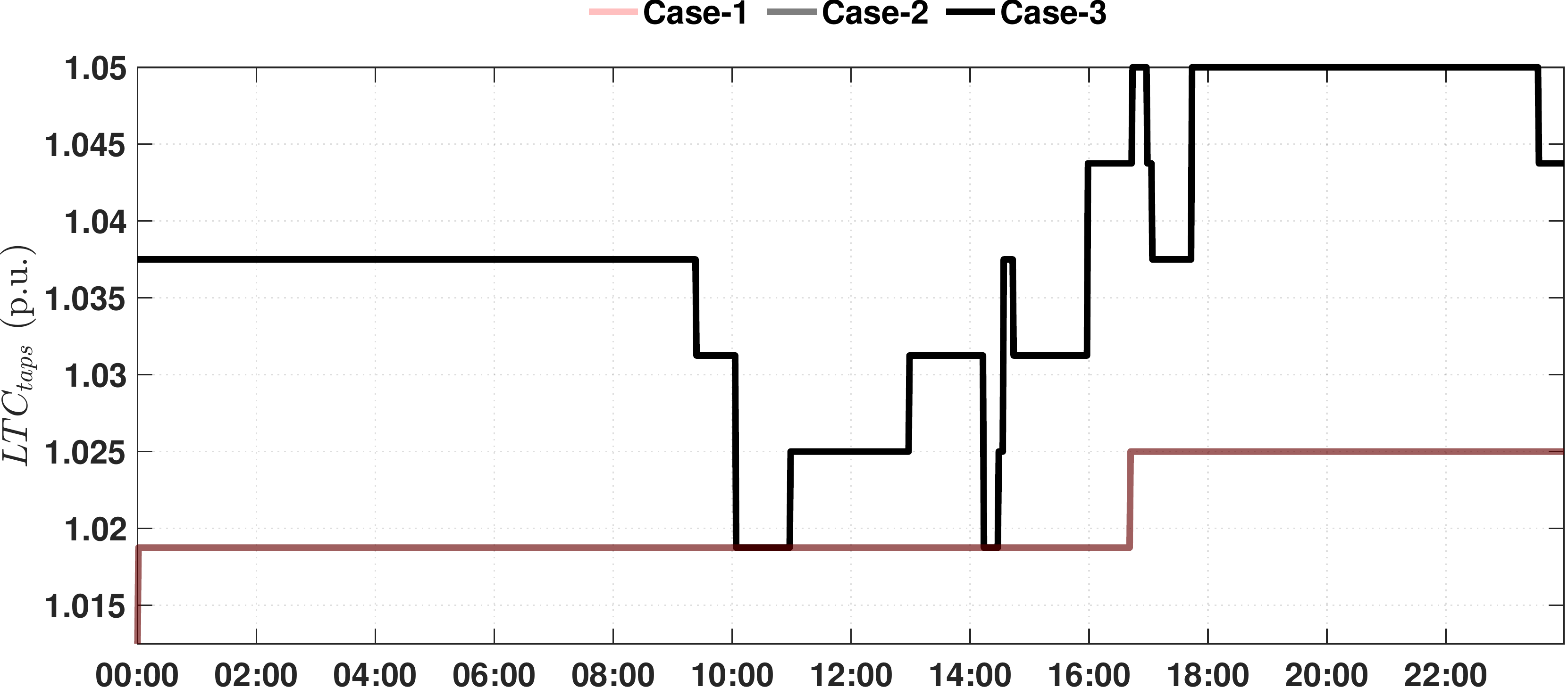}
  \caption{LTC tap change for High Load Cloudy PV condition}
  \label{fig:ltcplot}
\end{figure}

The voltage drop issue associated with absorbing reactive power during high load condition is mitigated by keeping the
LTC taps at maximum position as shown in fig. \ref{fig:ltcplot} compared to case-2 and case-1. This also highlights the
advantage of coordinated control between the devices.

\begin{figure}[htpb]
  \centering
  \includegraphics[width=0.48\textwidth]{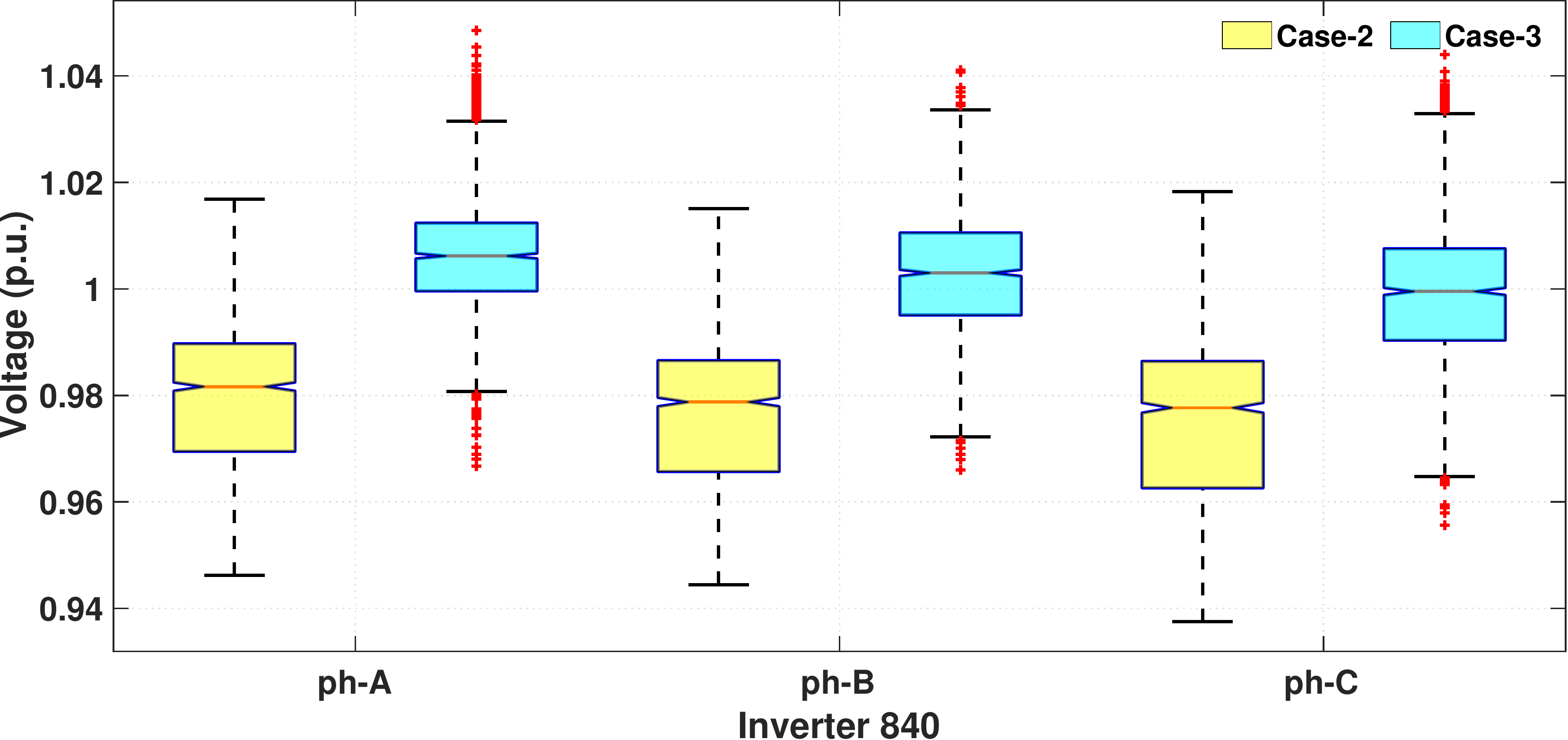}
  \caption{Comparison of 24h Voltage Profile at End Node 840 for High Load Cloudy PV condition}
  \label{fig:840V}
\end{figure}

Apart from the reduction in the reactive power usage, the main advantage of shifting the VVar-C is the smoothing of
system voltage profile. Figure \ref{fig:840V} shows the distribution of voltage at end node 840, which shows that
proposed case has a much tighter voltage profile compared to case-2. This is beneficial to utilities while applying
conservation voltage reduction (CVR).

\begin{figure}[htpb]
  \centering
  \includegraphics[width=0.48\textwidth]{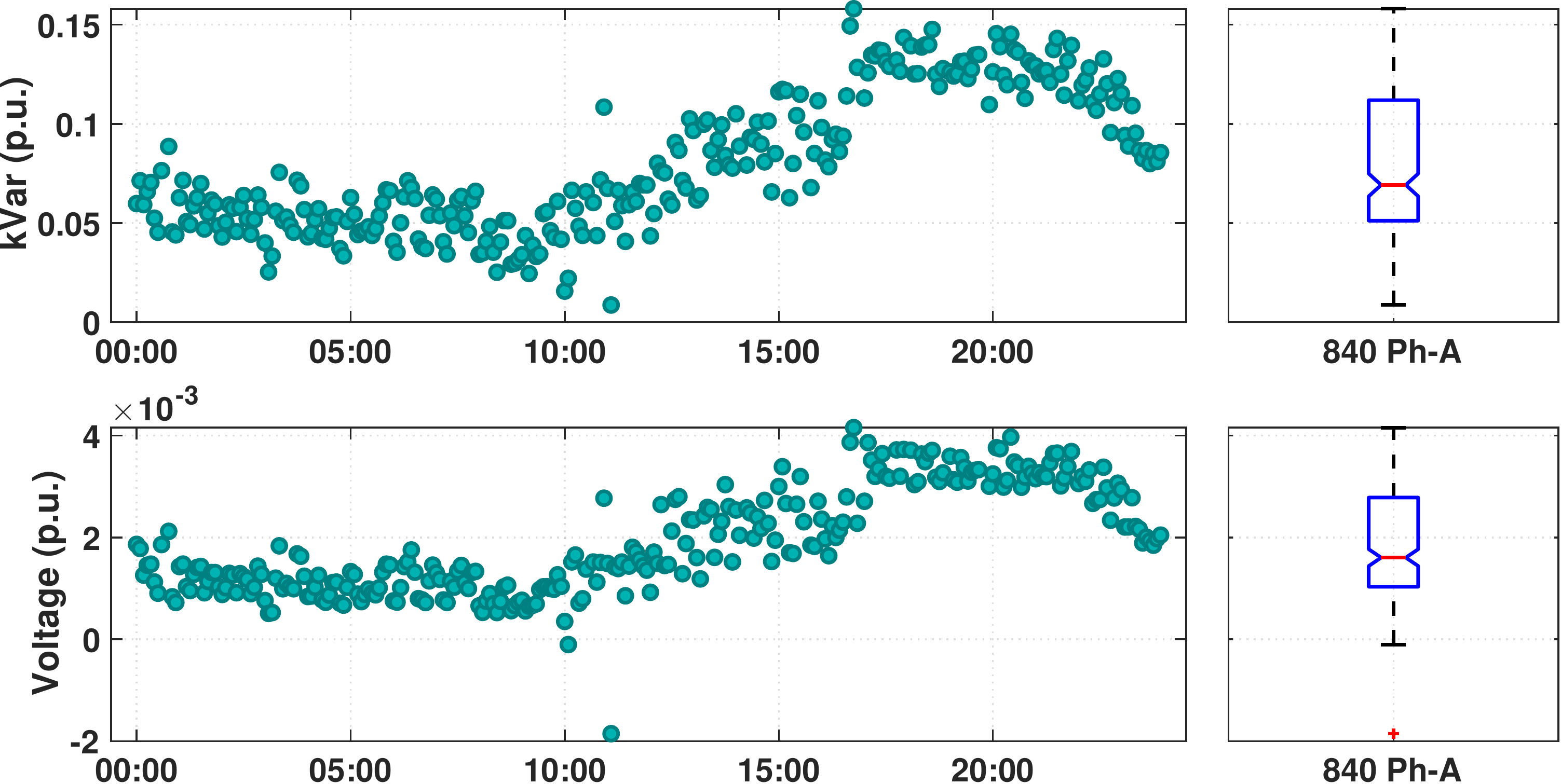}
  \caption{Error in reactive power injections (\textit{top}) and measured voltage (\textit{bottom}) vs optimal command
  at inverter 840 on phase-A at 5-min control intervals.}
  \label{fig:qerror}
\end{figure}

To validate the claim that by shifting the curve the equilibrium point of new curve is infact the optimal solution from
central VVO problem we show the error between the optimal var command against the var injection from new curve for same
operating points at 5-min intervals in fig. \ref{fig:qerror}. We can see that the error is very low about 7\% average
which is predominantly from the accumulation of inherent steady state error of Volt/VAR curves from all inverters in the
system which leads to a maximum voltage error of only 0.4\%.

\begin{figure}[htpb]
  \centering
  \includegraphics[width=0.48\textwidth]{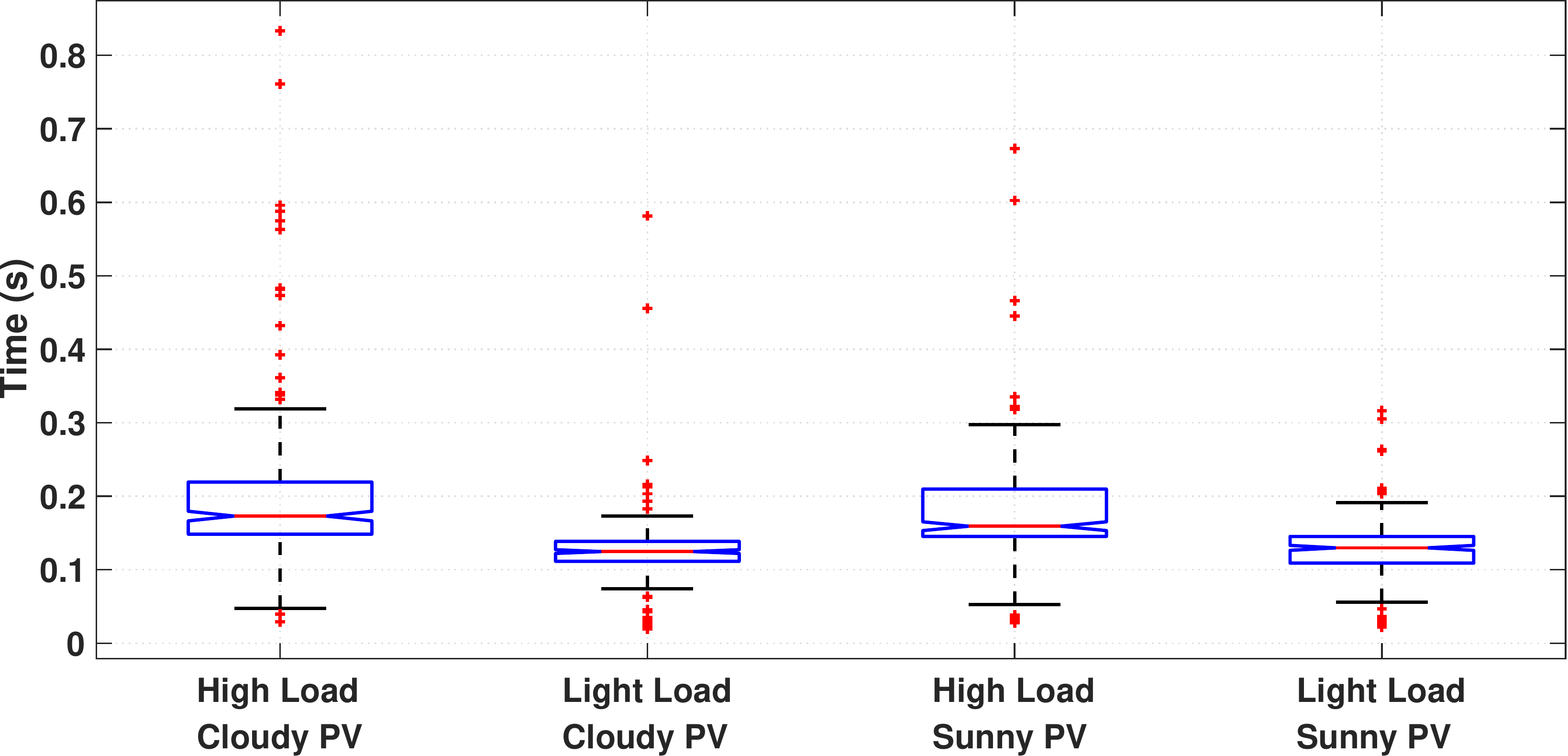}
  \caption{Time taken for Two-Stage Optimization Problem to Converge}
  \label{fig:time}
\end{figure}

Time taken for optimization problem to converge at every 5-minute interval is shown in fig. \ref{fig:time}. Under all
operating conditions the maximum time taken to converge is only 1 sec which indicates that the proposed algorithm can
easily be applied to larger systems with multiple voltage regulators and smart inverters while still dispatching with
5-min control interval.

\section{Conclusion}\label{sec:conclusion}
This paper proposes a new Volt/VAR optimization (VVO) scheme that coordinates the control of legacy devices like voltage
regulators and load tap-changers with smart inverters for real-time implementation. A new dispatching scheme for smart
inverters is proposed that utilizes the optimal var injections obtained from the coordinated VVO to shift the existing
volt/var curves in the inverters laterally thereby minimizing the communication requirement as well as reduce the
dependency on centralized VVO for voltage regulation. Simulation results on a test feeder shows that the new
VVO is superior to conventional volt/var control schemes based on local measurements in eliminating the voltage
violations, provide better voltage regulation, significant reduction in tap operation which will reduce the wear and
tear of the mechanical devices, and overall reduction in power loss in the circuit.

% \bibliography{/home/valli/Documents/MyLibrary.bib}
\bibliography{SecondDraft.bib}
% \bibliography{/home/valli/Documents/SecondBrain/MyLibrary.bib}
% \bibliography{/home/valli/Documents/MyLibrary.bib}

\section*{Appendix}
\subsection{Proof of Theorem-\ref{thm:curveshifting}}
\begin{proof}
Let $g(v, q) = 0$ denote the 3-phase power flow solution of an unbalanced distribution system. 

Let $q = f(v - v_{ref})$ denote the volt/var control function where $f\colon \mathbf{R}^n \to \mathbf{\Omega}$ denote the
set of individual inverter volt/var functions $f_i\colon \mathbf{R} \to \mathbf{\Omega_i}$. Note that $q \in \mathbf{\Omega}
= \prod_{i=1}^n \mathbf{\Omega_i}$ where, $\mathbf{\Omega}_i = \{q_i\mid \underline{q}_i \le q_i \le \overline{q}_i\}$
is determined by the inverter limits. Generally, $\underline{q}_i$ is 0 and  $\overline{q}_i$ is the available reactive
power determined by the current real power output and inverter size.

A point $(v^*, q^*)$ is an equilibirium point of control function $f$ if it satisfies the following equations
\cite{zhou2021a}:
\begin{equation}
  \begin{aligned}
    &g(v^*, q^*) = 0\\
    &q^* = f(v^* - v_{ref})
  \end{aligned}
\end{equation}
Given an optimal power flow solution $(v_g, q_g)$ which implies $g(v_g, q_g) = 0$. This point will not be an
equilibirium point of $f$ but we can still find a voltage $v_l$  corresponding to $q_g$ since $q_g \in \mathbf{\Omega}$
as the inverter limits are included as a constraint in optimal power flow problems. This implies $v_l = f^{-1}(q_g)$ or $q_g
= f(v_l - v_{ref})$, note that $f^{-1}$ exist since $f$ is non-increasing. $(v_l, q_g)$ will not be an equilibirium
point since it does not solve the power flow equations, $g(v_l, q_g) \neq 0$.

By shifting the function $f$ along the voltage axis by $v_g - v_l$ we have,
\begin{equation}
  \begin{aligned}
    q_g &= f(v_l - v_{ref} + v_g - v_l)\\
    q_g &= f(v_g - v_{ref})
  \end{aligned}
\end{equation}
which shows that $(v_g, q_g)$ is a solution of the shifted function $f(v - v_{ref} + v_g - v_l)$ and it is also the
equilibirium point of the system since it satisfies the power flow equations $g(v_g, q_g) = 0$. The direction of shift
will be implicitly taken care by the sign of $v_g - v_l$.

This shows that the an optimal inverter set point can be dispatched to inverters without modeling the volt/var curves in
the opitmal power flow problem which can become computationally intesive. There are no restrictions on the volt/var
curve $f$ and the optimal power flow problem, so any objective can be utilized and the volt/var function $f$ can be
continuously shifted by using the previous curve information.
\end{proof}

\vskip -2.5\baselineskip plus -1fil
\vspace{11pt}
\begin{IEEEbiography}[{\includegraphics[width=1in,height=1.25in,clip,keepaspectratio]{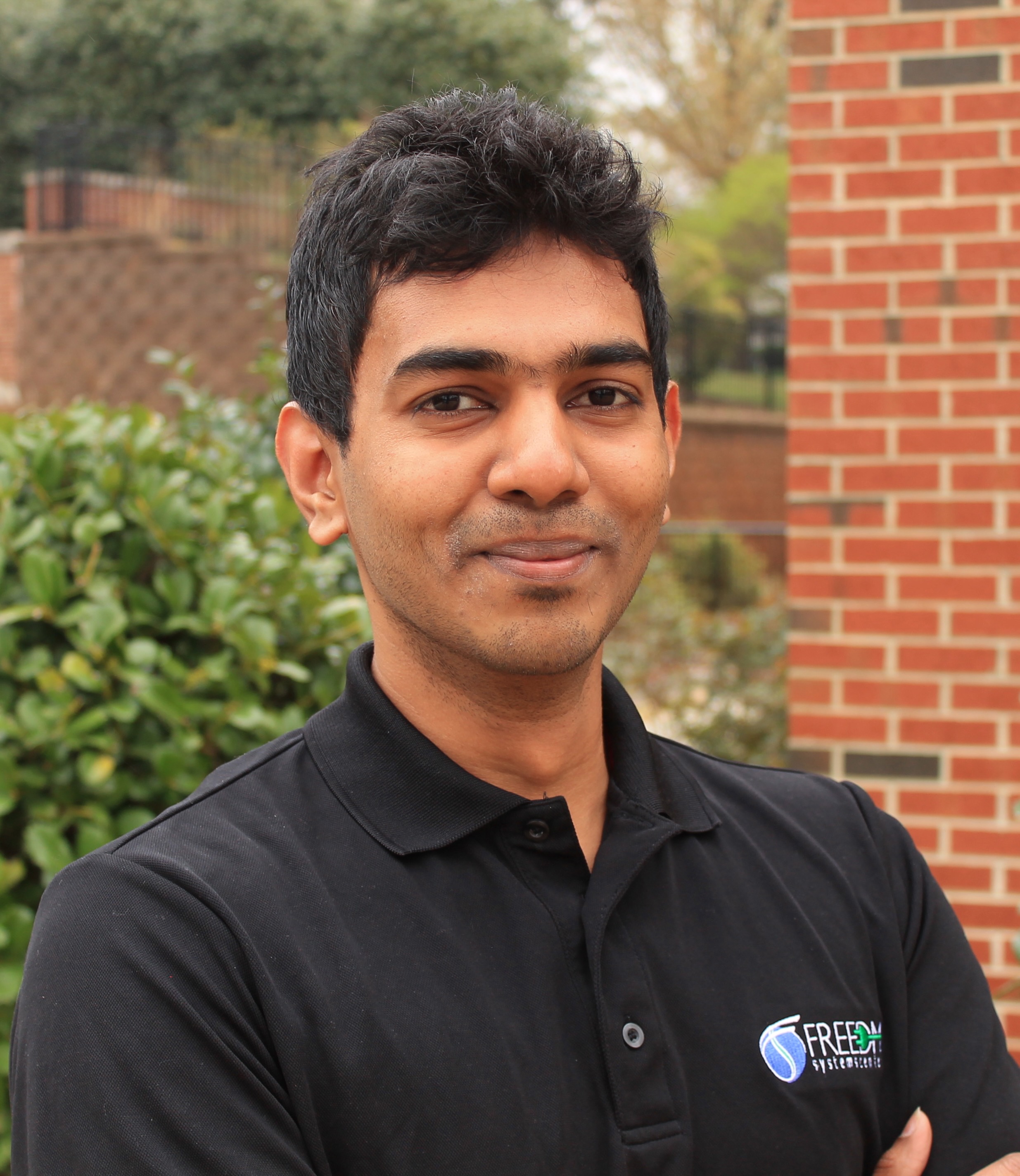}}]{Valliappan Muthukaruppan}
(S'17) received the B.Tech degree in electrical engineering from National Institute of
Technology, Trichy, India in 2014, and the M.S. degree in electrical engineering from North Carolina State University,
Raleigh, NC, USA in 2018. He is currently working towards the Ph.D. degree in electrical engineering at North Carolina
State University.

His research interests include application of optimization, data science, and machine learning techniques to control and
operate smart distribution networks with high renewable penetration.
\end{IEEEbiography}

\vskip -2\baselineskip plus -1fil
\vspace{11pt}
\begin{IEEEbiography}[{\includegraphics[width=2in,height=1.25in,clip,keepaspectratio]{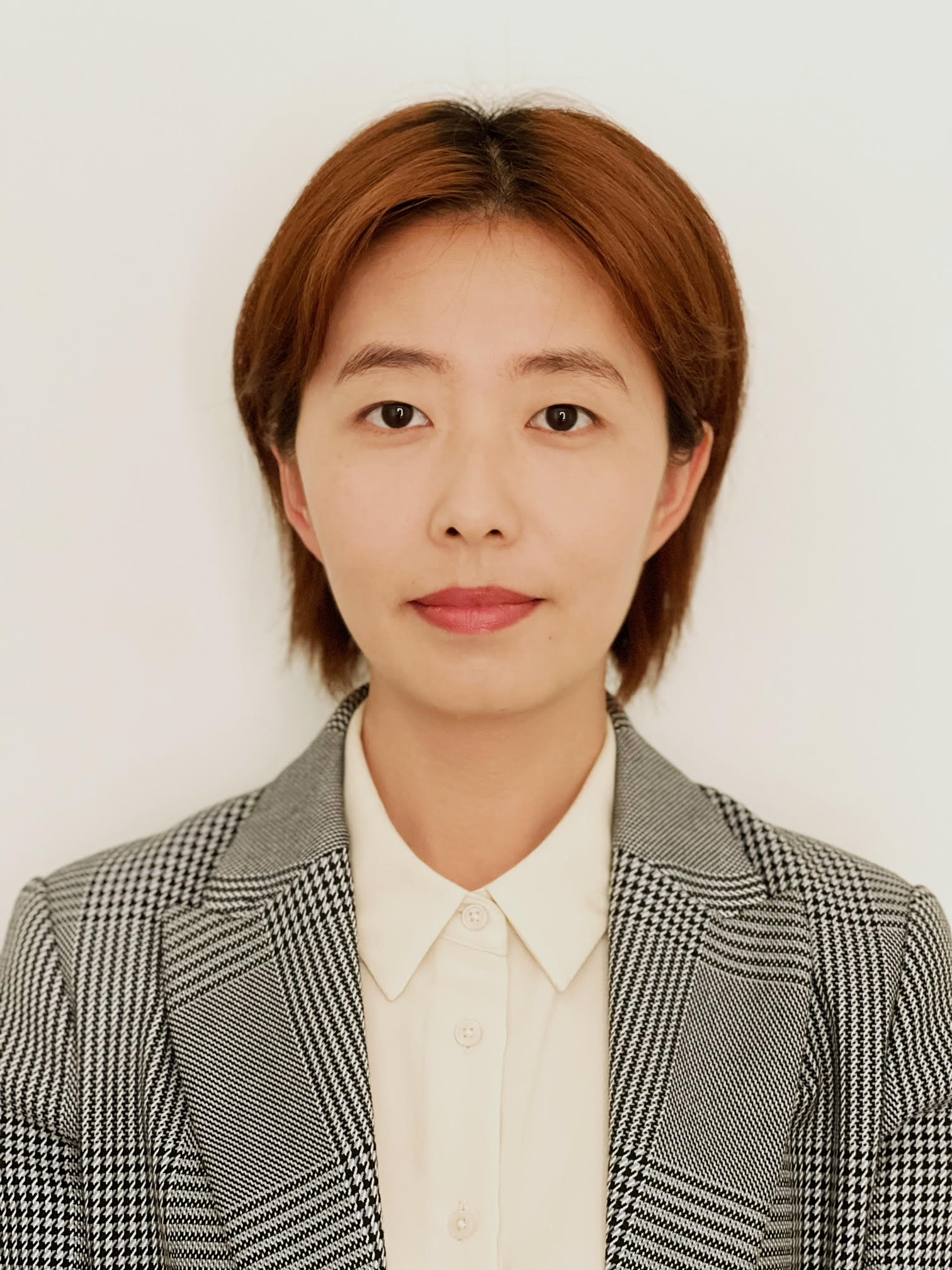}}]{Yue Shi}
(S'12-M'18-SM'19) received the B.E. degree in electrical engineering from Tianjin University, Tianjin, China, in 2012, the M.S.
and Ph.D. degree in electrical engineering from North Carolina State University, Raleigh, NC, USA in 2014 and 2018.

Her research interests include Volt/Var optimization, electric utility asset investment optimization, and impacts of
renewable energy and electric vehicles on power distribution systems.
\end{IEEEbiography}

\vskip -2\baselineskip plus -1fil
\vspace{11pt}
\begin{IEEEbiography}[{\includegraphics[width=1in,height=1.25in,clip,keepaspectratio]{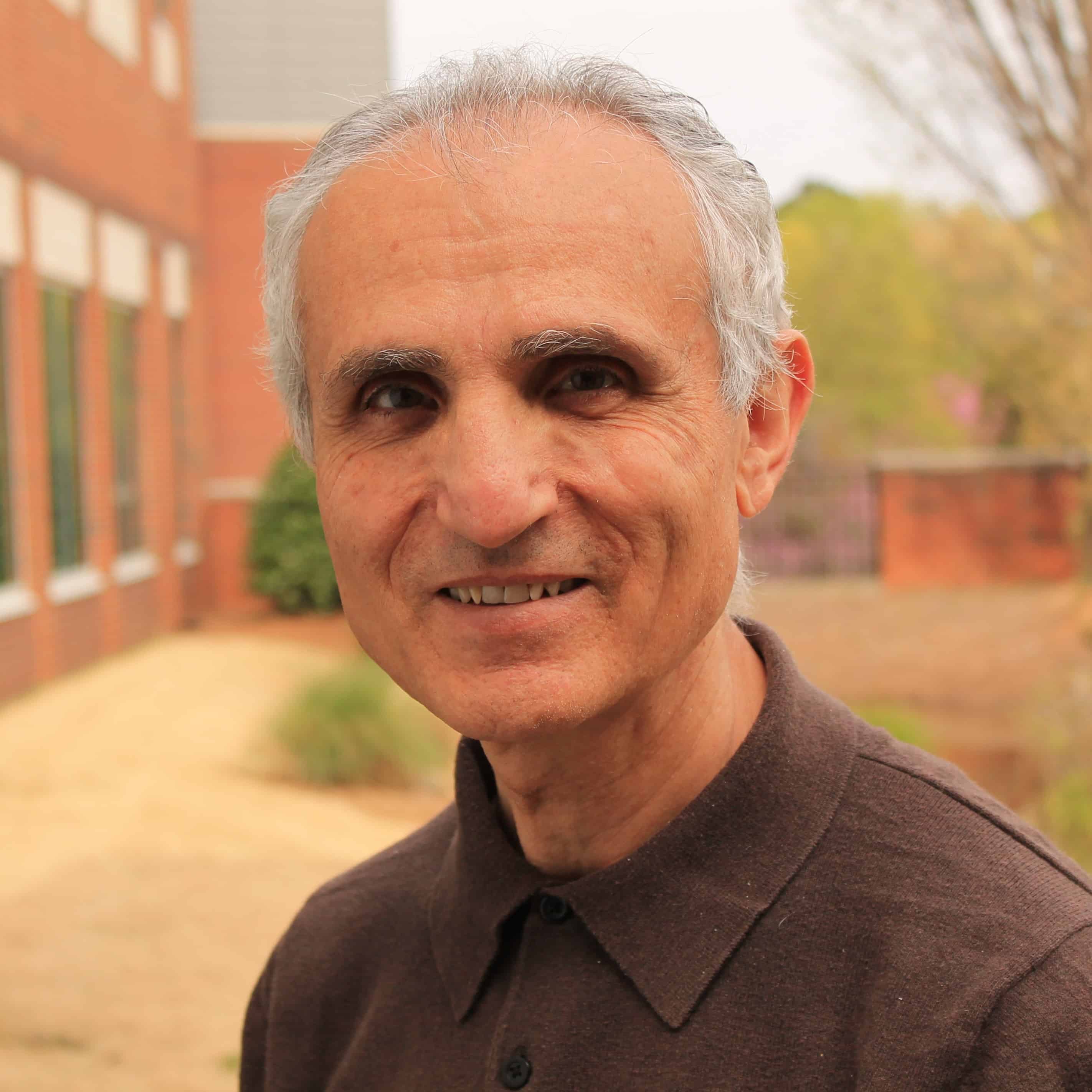}}]{Mesut E.
Baran}
(S'87-M'88-SM'05-F'11) received the B.S. and M.S. degree from Middle East Technical University, Ankara, Turkey, and the
Ph.D. degree from the University of California, Berkeley, CA, USA, all in electrical engineering. He is currently
a Professor with North Carolina State University, Raleigh, NC. 

His research interests include distribution and transmission system analysis and control, integration of renewable
energy resources, and utility applications of power electronics based devices. He is currently a member of the FREEDM
Systems Center with North Carolina State University focusing on both research and education aspects of renewable
electric energy systems and their integration in to the electric power distribution system.
\end{IEEEbiography}
\iffalse
\newpage

\section{Biography Section}
If you have an EPS/PDF photo (graphicx package needed), extra braces are
 needed around the contents of the optional argument to biography to prevent
 the LaTeX parser from getting confused when it sees the complicated
 $\backslash${\tt{includegraphics}} command within an optional argument. (You can create
 your own custom macro containing the $\backslash${\tt{includegraphics}} command to make things
 simpler here.)
 
\vspace{11pt}

\bf{If you include a photo:}\vspace{-33pt}
\begin{IEEEbiography}[{\includegraphics[width=1in,height=1.25in,clip,keepaspectratio]{fig1}}]{Michael Shell}
Use $\backslash${\tt{begin\{IEEEbiography\}}} and then for the 1st argument use $\backslash${\tt{includegraphics}} to declare and link the author photo.
Use the author name as the 3rd argument followed by the biography text.
\end{IEEEbiography}

\vspace{11pt}

\bf{If you will not include a photo:}\vspace{-33pt}
\begin{IEEEbiographynophoto}{John Doe}
Use $\backslash${\tt{begin\{IEEEbiographynophoto\}}} and the author name as the argument followed by the biography text.
\end{IEEEbiographynophoto}

\vfill
\fi

\end{document}